\documentclass[twocolumn,aps,pra,longbibliography,superscriptaddress]{revtex4-1}
\usepackage[latin9]{inputenc}
\usepackage{color}
\usepackage{textcomp}
\usepackage{amsmath}
\usepackage{amssymb}
\usepackage{graphicx}
\usepackage{esint,hyperref}

\makeatletter
\usepackage{amsfonts}
\usepackage{bbold}
\usepackage{natbib}

\pdfoutput=1
\newcommand{\ket}[1]{|#1 \rangle}
\newcommand{\bra}[1]{\langle#1 |}

\newcommand{\lr}[1]{\left( #1 \right)}

\newcommand{\mean}[1]{\langle #1 \rangle}
\newcommand{\no}{\nonumber}
\newcommand{\tr}[1]{\mathrm{tr}\left\{#1\right\}}
\newcommand{\ptr}[2]{\mathrm{tr}_{#1}\left\{#2\right\}}

\newcommand{\be}{\begin{equation}}
\newcommand{\ee}{\end{equation}}

\begin{document}

\title{Measurement Protocol for the Entanglement Spectrum of Cold Atoms}

\author{Hannes Pichler}
 \thanks{hannes.pichler@cfa.harvard.edu} 
\affiliation{ITAMP, Harvard-Smithsonian Center for Astrophysics, Cambridge, Massachusetts 02138, USA}
\affiliation{Physics Department, Harvard University, Cambridge, Massachusetts 02138, USA}
\affiliation{Institute for Quantum Optics and Quantum Information of the Austrian
Academy of Sciences, A-6020 Innsbruck, Austria}

\author{Guanyu Zhu}
 \thanks{gzhu123@umd.edu} 
\affiliation{Joint Quantum Institute, NIST/University of Maryland, College Park, Maryland 20742, USA}

\author{Alireza Seif}
\affiliation{Joint Quantum Institute, NIST/University of Maryland, College Park, Maryland 20742, USA}

\author{Peter Zoller}
\affiliation{Institute for Quantum Optics and Quantum Information of the Austrian
Academy of Sciences, A-6020 Innsbruck, Austria}
\affiliation{Institute for Theoretical Physics, University of Innsbruck, A-6020
Innsbruck, Austria}

\author{Mohammad Hafezi}
\affiliation{Joint Quantum Institute, NIST/University of Maryland, College Park, Maryland 20742, USA}
\affiliation{Kavli Institute for Theoretical Physics, Santa Barbara, California 93106, USA}
\affiliation{Department of Electrical and Computer Engineering and Institute for Research in Electronics and Applied Physics,
University of Maryland, College Park, Maryland 20742, USA}

\begin{abstract} 
Entanglement, and, in particular the entanglement spectrum, plays a major role in characterizing many-body quantum systems. While there has been a surge of theoretical works on the subject, no experimental measurement has been performed to date because of the lack of an implementable measurement scheme. Here, we propose a measurement protocol to access the entanglement spectrum of many-body states in experiments with cold atoms in optical lattices. Our scheme effectively performs a Ramsey spectroscopy of the entanglement Hamiltonian and is based on the ability to produce several copies of the state under investigation together with the possibility to perform a global swap gate between two copies conditioned on the state of an auxiliary qubit.  We show how the required conditional swap gate can be implemented with cold atoms, either by using Rydberg interactions or coupling the atoms to a cavity mode. We illustrate these ideas on a simple (extended) Bose-Hubbard model where such a measurement protocol reveals topological features of the Haldane phase. 
\end{abstract}

\date{\today}

\maketitle

\section{Introduction}
Nowadays, entanglement is a central concept  in many branches of quantum physics ranging from quantum information \cite{Nielsen:2011:QCQ:1972505} to condensed matter theory \cite{Amico:2008en, Eisert:2010uz, Laflorencie:2015vc}  and high-energy physics  \cite{Srednicki:1993dl}. Given a many-body quantum state $\rho$, a fundamental quantity in characterizing the bipartite entanglement between two subsystems $\mathcal{A}$ and $\mathcal{B}$ is the corresponding reduced density operator of a partition, $\rho_{\mathcal{A}}\equiv\ptr{{\mathcal{B}}}{\rho}$  \cite{Horodecki:2009gb}. In particular, the entanglement spectrum (ES), the spectrum of $\rho_{\mathcal{A}}$, denoted as $\sigma(\rho_{\mathcal{A}})$, has proven to be a powerful theoretical tool to analyze entanglement properties in a quantum information context \cite{Nielsen:1999ct,Nielsen:2001gh,Horodecki:2009gb} and has more recently also attracted much interest in condensed matter physics to characterize many-body quantum states. 
For example, as first pointed out by Haldane and Li \cite{Li:2008cg}, the ES can serve as fingerprint of topological order since it mimics the excitation spectrum of chiral edge modes \cite{Thomale:2010it,  Cirac:2011jq,  Chandran:2011ex, Qi:2012jv, Pollmann:2010ih,  Schuch:2013hf}. In this context, the central spirit lies in the so-called bulk-edge correspondence of the ES \cite{Chandran:2011ex, Qi:2012jv}, a manifestation of the holographic principle \cite{Srednicki:1993dl, Eisert:2010uz,Cirac:2011jq}.    Moreover, the importance of the ES has been discussed in the context of tensor networks  \cite{Cirac:2011jq, Schuch:2013hf}, quantum criticality \cite{Calabrese:2008cm}, symmetry-breaking phases \cite{Cirac:2011jq, Metlitski:2011, Alba:2012bt}, and, most recently, many-body localization \cite{Yang:2015ga, Berkovits:2015, Regnault:2016} and eigenstate thermalization \cite{GroverETH:2015}.
At the same time, it has been pointed out in the literature \cite{Chandran:2014ch} that the ES can contain nonuniversal features requiring caution in using it as a tool to locate phase transitions, in particular, for symmetry-breaking phases.  Given the enormous interest in the ES as an important theoretical concept and powerful numerical tool, an outstanding challenge, however, is the direct experimental measurement of the ES in many-body systems, where a full state tomography is prohibitively expensive or even impossible. At the same time, the rapid development of cold-atom experiments \cite{Zeiher:2015gw,Parsons:2015vu,Cheuk:2015jr,Kennedy:2015dw,Baier:2016ga,Jotzu:2014kz} has offered unique opportunities to access quantities related to entanglement in a many-body system \cite{MouraAlves:2004jv,Daley:2012ud,Abanin:2012bj,Hauke:2016ht}. A remarkable achievement in this context is the recent measurement of the second Renyi entropy $(\mathcal{S}_{2}=-\log \rm{Tr} \rho^2_\mathcal{A})$, i.e., the purity of the reduced density operator in cold-atom experiments in optical lattices \cite{Islam:2015cm, Kaufman:2139247} following the theoretical proposals in Refs.~\cite{MouraAlves:2004jv,Daley:2012ud}.

Motivated by these recent developments, we present a protocol on how the ES can be directly measured in a quantum simulator. In analogy to a many-body Ramsey interferometry \cite{Ekert:2002uu,Muller:2009fp,Knap:2013jy,Zeiher:2016td}, where the evolution of a many-body state is conditioned on an ancilla system,  we develop a scheme where  the conditional  evolution of the many-body system is determined by a copy of the density operator, acting as the Hamiltonian \cite{Lloyd:2014gc}.  This is achieved by a sequence of global swap operations between two copies of the system, controlled by the ancilla. The Fourier transform of the ancilla's Ramsey signal reveals the ES and the corresponding degeneracies. We discuss the physical implementation of this scheme in a cold-atom setup, relying on a combination of single-site-resolved addressing techniques in optical lattices  \cite{Preiss1229,Fukuhara:2015ec} and dispersive interactions with an ancilla atom via the Rydberg blockade mechanism \cite{Saffman:2010ky,Comparat:2010cb,Hofmann2014,Maller:2015is}.   We illustrate our protocol with an extended Bose-Hubbard model containing the Haldane phase \cite{HaldaneConjecture1983, Affleck:1987jy}.  Our detection of the degeneracies of the ``ground state'' in ES and the entanglement gap serves as a signature \cite{Pollmann:2010ih} of the symmetry-protected topological phase \cite{ChenScience}.   

\begin{figure}[t]
\includegraphics[width=0.45\textwidth]{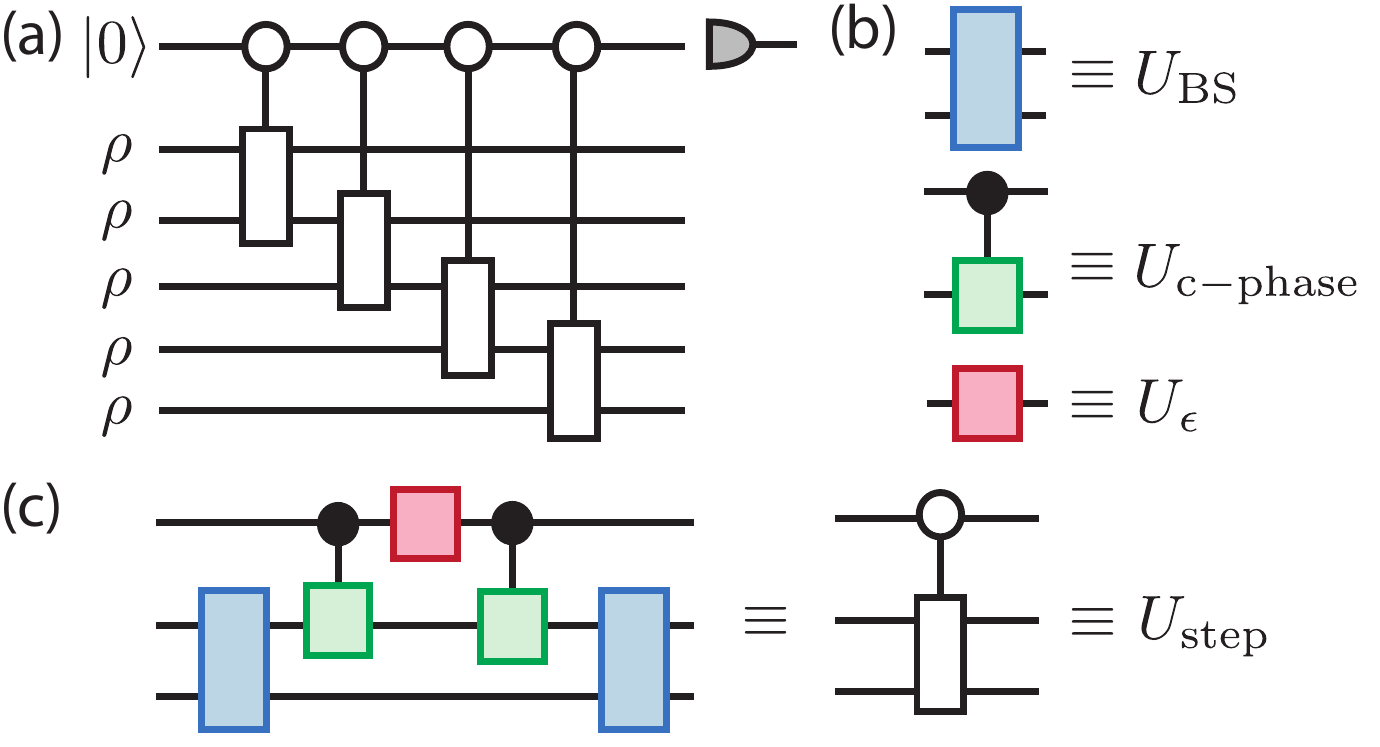}
\caption{(a) Circuit representation of the protocol to determine the spectrum of a density operator $\rho$. It consists of $n$ stroboscopic steps (operations) that each involve an ancilla system and two copies of the state under investigation. 
(b) Each step can be constructed from three basic operations: tunnel coupling between neighboring copies (blue), a controlled (dispersive) phase shift for atoms in the lattice based on the state of the ancilla system (green), and rotations of the ancilla qubit (red). For a description of these operations, we refer to the main text [Eqs.~\eqref{BS},~\eqref{cPhasegate}, and \eqref{Usinglequbit}; see also Fig. \ref{fig2}]. (c) Construction of the stroboscopic step $U_{\rm step}$ from these elementary operations. Note that it differs from the one in Eq.~\eqref{eq2} by an additional swap, $U_{\rm step}=SU_{\rm step}'$. This additional swap ensures that all processes involve only neighboring copies (see Appendix \ref{App:Protocol}).
}
\label{fig1}
\end{figure}

\section{Protocol} 
To outline the key idea we consider two quantum systems, labeled $1$ and $2$, in the quantum states $\rho_1$ and $\rho_2$, respectively. Here, the Hilbert spaces for system $1$ and system $2$ are assumed to be isomorphic.
The basic building block of our protocol relies on the identity \cite{Lloyd:2014gc}
\begin{align}\label{basic_step}
\textrm{tr}_2\{e^{-i\epsilon S}\rho_1\otimes \rho_2\, e^{i\epsilon S}\}=\rho_1-i\epsilon[\rho_2,\rho_1]+\mathcal{O}(\epsilon^2),
\end{align}
where $S$ is the swap operator that interchanges the quantum state of systems $1$ and $2$: $S\ket{\psi_1}\otimes\ket{\psi_2}=\ket{\psi_2}\otimes\ket{\psi_1}$. The right-hand side of Eq.~\eqref{basic_step} describes a coherent evolution of system 1, which is not generated by a Hamiltonian but instead by the density operator of the second system, $\rho_2$. Below, we show how one can reduce the required unitary $U_S(\epsilon)=e^{-i\epsilon S}$ to a set of simpler operations that can be implemented in state-of-the-art experiments with cold atoms.

The central idea is now to repeatedly perform the operation $U_S(\epsilon)$ and obtain a stroboscopic evolution of $\rho_1$ with $\rho_2$ according to  Eq.~\eqref{basic_step}, using a new copy of $\rho_2$ in each step. Therefore, after $n$ steps, we obtain the map $\rho_1\rightarrow \mathcal{E}_{\rho_2}(\rho_1)=e^{-i n\epsilon \rho_2}\rho_1 e^{i n\epsilon \rho_2}$. Thus, $\rho_2$ takes the role of a Hamiltonian for system $1$, which evolves ``for a time'' $t_n=n\epsilon$.  Monitoring this evolution allows us to access the spectral properties of $\rho_2$, e.g., via quantum phase estimation using an ancillary quantum computer \cite{Lloyd:2014gc}. Here we use a simple Ramsey technique instead and employ an ancillary system (with basis states $\ket{0}$ and $\ket{1}$) to control the application of $U_S(\epsilon)$ [cf. Fig.~\ref{fig1}(a)]:
\begin{align}\label{eq2}
U_{\rm step}'=\ket{-}\bra{-}\otimes U_S(-\epsilon)+\ket{+}\bra{+}\otimes U_S(+\epsilon),
\end{align} 
where $\ket{\pm}=(\ket{0}\pm\ket{1})/\sqrt{2}$. For an ancilla initially prepared in the state $\ket{0}$, the measurement of the operator $Z=\ket{0}\bra{0}-\ket{1}\bra{1}$ after $n$ such controlled stroboscopic steps gives $\mean{Z}_n=\frac{1}{2}\textrm{tr}\{e^{-i2t_n \rho_2}\rho_1+\rho_1 e^{i2t_n\rho_2}\}$. This expression is valid for small time steps, such that $t_n\epsilon\ll1$. For the general expression, we refer to Appendix \ref{App:Protocol}. In particular, for $\rho_1=\rho_2\equiv\rho$, one gets
\begin{align}\label{Ramsey}
\mean{Z}_n=\textrm{tr}\{\rho\cos(2t_n \rho)\}=\sum_{\alpha}\lambda_\alpha\cos(2t_n \lambda_\alpha).
\end{align}
The set of eigenvalues of $\rho$, $\{\lambda_\alpha\}$, can thus be extracted by a simple Fourier transform of the measurement signal for different $n$. Note that the choice of $\rho_1=\rho_2=\rho$ is not fundamental, but it is a natural one for an experimental implementation and renders the protocol sensitive to the largest eigenvalues. 

The basic building block of this protocol is the unitary $U_S(\epsilon)=e^{-i\epsilon S}$ on the joint system of $1$ and $2$. We note that the recently implemented schemes to measure the Renyi entropy of cold atoms \cite{Islam:2015cm,Kaufman:2139247} require a measurement of the expectation value of the swap operator $S$. For a many-body system, $S$ can be decomposed into a product of local swap operators, such that a measurement is possible by local operations only \cite{Daley:2012ud,Abanin:2012bj,Pichler:2013bs}. Here we aim for a more ambitious goal since we want to apply a unitary that is generated by the global swap operator. This is a nontrivial task since $S$, and, therefore, $U_S$ are highly non-local. Remarkably, the protocol outlined above can nevertheless be implemented with operations relying only on experimental tools already available in cold atom experiments as discussed below, such as controlled tunneling between neighboring lattice sites \cite{Islam:2015cm}, local addressability of individual sites  \cite{Preiss1229,Fukuhara:2015ec} and dispersive interactions based on the Rydberg blockade mechanism \cite{Saffman:2010ky}.

\section{Cold-atom implementation} \label{Implementation}
While the protocol discussed above is completely general, we now present an implementation thereof in experiments with cold atoms in optical lattices. For concreteness, we consider bosons in a one-dimensional optical lattice described by a Bose-Hubbard model \cite{Jaksch:1998to}, but the protocol equally applies to two-dimensional systems. We are interested in the entanglement (spectrum) between different lattice sites of an arbitrary state $\rho$, i.e., the  {\em motional} degrees of freedom of the system. A realization of the above Ramsey interferometer with cold atoms consists of the following steps:

\begin{figure}[t]
\includegraphics[width=0.45\textwidth]{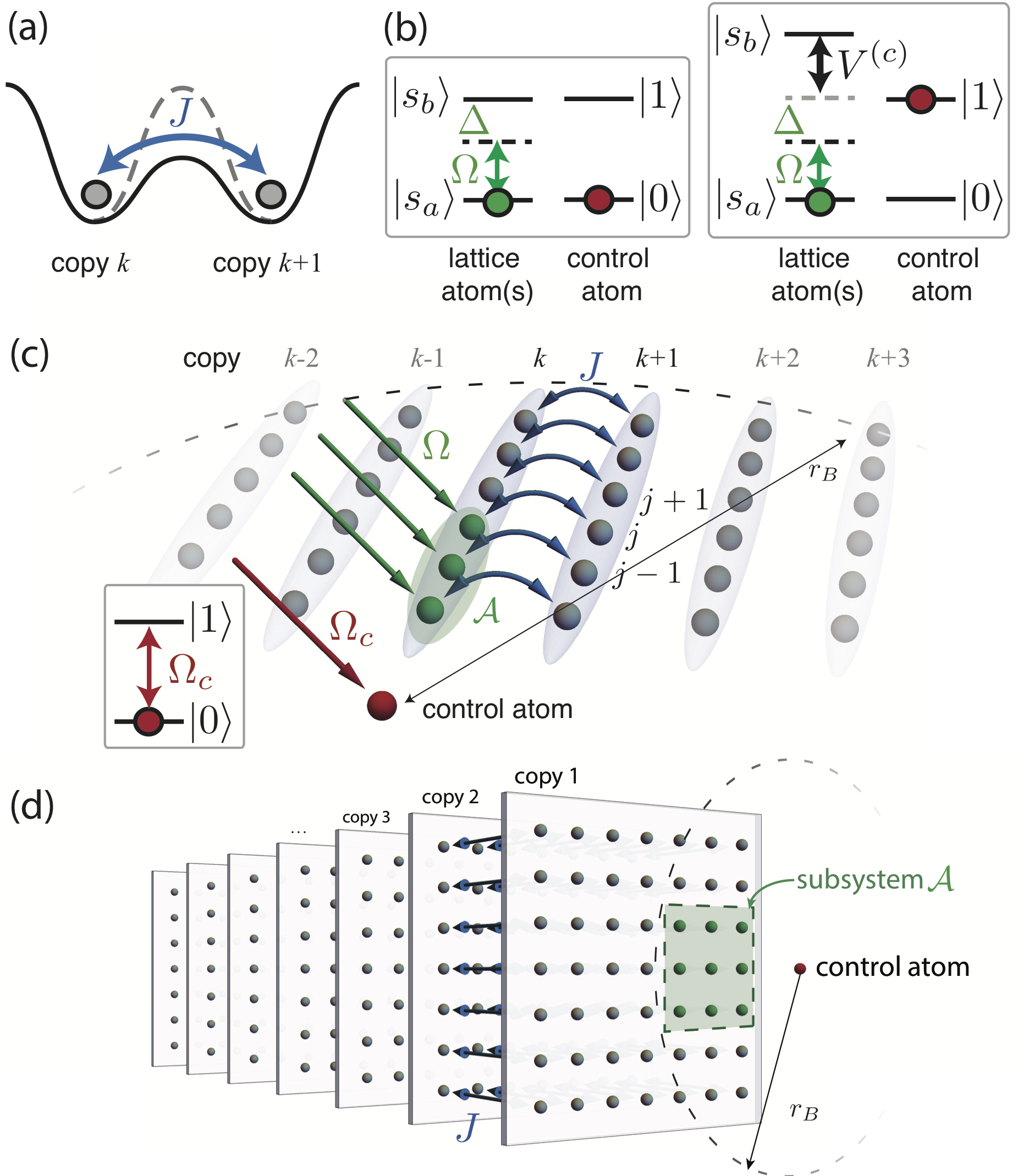}
\caption{Realization of the elementary operations of the circuit in Fig.~\ref{fig1} with cold atoms. (a) The beam splitter  $U_{\rm BS}$ [cf.~Eq.~\eqref{BS}] can be realized by lowering the barrier between two copies in order to allow for tunneling between them. (b) To realize the controlled phase shift, an off-resonant laser to a Rydberg state $\ket{s_b}$ is focused on the lattice sites in subsystem $\mathcal{A}$ of copy $k$. If the control atom is in the state $\ket{0}$, the ac-Stark shift leads to an acquired phase (left); if it is in the Rydberg state $\ket{1}$, the Rydberg blockade mechanism leads to a shift of the state $\ket{s_b}$, suppressing the ac-Stark shift (right). (c) Illustration of the basic operations in step $k$ to determine the spectrum of the reduced density operator of only the first three lattice sites. Note that the blockade radius $r_B$ exceeds the size of the lattice. The rotations of the control atom $U_{\epsilon}$ correspond to Rabi pulses on a Rydberg transition. (d) Analogous setup for 2D systems. Several copies are created in different layers of a 3D lattice, with the ancillary atom trapped nearby.  All steps in the protocol are the same as in the 1D case. We note that the subsystem $\mathcal{A}$ does not need to be contiguous. 
}
\label{fig2}
\end{figure}

(i) Preparation: $n+1$ identical copies of the many-body state $\rho$ are prepared. Such copies are naturally realized in experiments with cold atoms in optical lattices \cite{Islam:2015cm}, e.g,.~in parallel one-dimensional (1D) tubes. Once the states are produced, we freeze the motion along each lattice for the duration of the whole protocol by rapidly increasing the lattice depth and turning off interactions between the atoms, e.g., via Feshbach resonances \cite{Chin:2010vi}. 
In addition, an atom, whose internal states represent the ancilla qubit, is trapped in a separate lattice site, close to one of the tubes  (Fig.~\ref{fig2}).  We initialize the ancilla atom in a stable state, representing $\ket{0}$, and use a highly excited Rydberg state to represent $\ket{1}$. 

(ii) Stroboscopic steps: A single run of the protocol consists of $n$ stroboscopic steps. The $k$th step  ($k=1,\dots, n$) involves only the ancilla atom and the atoms in the two adjacent lattices $k$ and $k+1$ (Fig.~\ref{fig1}). We label the sites in each lattice by $j=1,\dots, M$, and the bosonic annihilation operator in the $k$th copy by $a_{j,k}$. In each single stroboscopic step, three types of operations are performed: (a) A tunnel coupling is induced between all the sites in lattice $k$ and $k+1$ [Fig.~\ref{fig2}(a)]. By lowering the potential barrier between the two lattices, e.g., using a superlattice, the atoms are allowed to tunnel between the two copies, as described by the Hamiltonian: $H_{\rm BS}=-J\sum_{j=1}^M (a_{j,k}^\dag a_{j,k+1}+a_{j,k+1}^\dag a_{j,k})$ \cite{Islam:2015cm}. After the time $t_{\rm BS}=\pi/(4J)$ the potential is ramped up again, realizing the so-called ``beam splitter'', 
 \begin{align}\label{BS}
U_{\rm BS}=\exp\bigg(i\frac{\pi}{4}\sum_{j} (a_{j,k}^\dag a_{j,k+1}+a_{j,k+1}^\dag a_{j,k})\bigg).
\end{align} 
(b) The second type of process involves a controlled phase shift \begin{align}\label{cPhasegate}U_{\rm c-phase}=\ket{1}\bra{1}\otimes\mathbb{1}+\ket{0}\bra{0}\otimes (-1)^{\sum_{j\in\mathcal{A}} a_{j,k}^\dag a_{j,k}},\end{align} where a $\pi$ phase is acquired by the atoms in sites $j\in \mathcal{A}$ of lattice $k$ depending on the state of the ancilla atom.
To achieve this, we make use of the internal structure of the atoms in the two chains, which so far were treated as structureless bosons in a stable internal state $\ket{s_a}$. In order to realize the controlled phase shift, we couple the atoms in  subsystem $\mathcal{A}$ of lattice $k$ to a Rydberg state $\ket{s_b}$ using a laser with Rabi frequency $\Omega$, which is detuned by $\Delta$ from the $\ket{s_a}\leftrightarrow \ket{s_b}$ transition. The corresponding ac-Stark shift gives the required $\pi$-phase shift. However, if the control atom is in the Rydberg state $\ket{1}$, the Rydberg blockade mechanism suppresses this process [see Fig.~\ref{fig2}(b)].
On a more formal level, the corresponding Hamiltonian for this process reads
\begin{align}
&H=\sum_{j\in\mathcal{A}}\lr{ -\Delta b_{j,k}^\dag b_{j,k}+\Omega (b_{j,k}^\dag a_{j,k}+a_{j,k}^\dag b_{j,k})}\no\\
&+\sum_{j} V_j^{(c)} \ket{1}\bra{1}\otimes b_{j,k}^\dag b_{j,k}+\sum_{j,l} V_{j,l}^{(b)} b_{j,k}^\dag b_{l,k}^\dag b_{l,k}b_{j,k}.\label{Rydbergintercations}
\end{align} 
Here, we denote the bosonic annihilation operator for an atom in the Rydberg state $\ket{s_b}$ on site $j$ in the $k$th copy by $b_{j,k}$. The second line of Eq.~\eqref{Rydbergintercations} contains the Rydberg-Rydberg interactions of an atom on site $j$ in the state $\ket{s_b}$ with the control atom in the Rydberg state $\ket{1}$ (with strength $V_j^{(c)}$), as well as interactions between Rydberg atoms in the lattice (with strength $V_{j,l}^{(b)}$).
In the fully blocked regime, for $|V_j^{(c)}|\gg |\Delta|\gg |\Omega|\sqrt{N_{\mathcal{A}}}$ for all $j\in \mathcal{A}$ (where $N_{\mathcal{A}}$ is the number of atoms in $\mathcal{A}$), one can adiabatically eliminate the state $\ket{s_b}$ and obtain, in second-order perturbation theory, the effective Hamiltonian, 
$H_{\rm eff}=\sum_{j\in\mathcal{A}} \frac{\Omega^2}{\Delta} \ket{0}\bra{0} \otimes a_{j,k}^\dag a_{j,k}.$
Note that the distance dependence of the Rydberg interactions ($V_j^{(c)}$) as well as interactions between lattice atoms ($V_{j,l}^{(b)}$) drop out in this regime. 
By applying $H_{\rm eff}$ for a time $t_{\rm phase}=\pi\Delta/\Omega^2$, one obtains the unitary given in Eq.~\eqref{cPhasegate}.
Such Rydberg-blockade-based gates have already been studied and demonstrated experimentally \cite{Maller:2015is,Saffman:2010ky,Zeiher:2016td}.
We point out that the AC-Stark laser field has to be applied in a site-resolved way  \cite{Preiss1229,Fukuhara:2015ec} only to the sites in $\mathcal{A}$ of copy $k$. 
(c) The third operation is a simple single-qubit rotation,
\begin{align}\label{Usinglequbit}
U_\epsilon&=\exp(-i\epsilon(\ket{0}\bra{1}+\ket{1}\bra{0})),
\end{align}
as realized by addressing the control atom with a resonant laser of Rabi frequency $\Omega_c$ for a time $t_\epsilon=\epsilon/\Omega_c$ [Fig.~\ref{fig2}(c)].
The rotation angle $\epsilon\ll1$ determines the stroboscopic step size.

Combining these operations one obtains a single stroboscopic step by $U_{\rm step}=U_{\rm BS}U_{\rm c-phase}U_{\epsilon}U_{\rm c-phase}U_{\rm BS}$ [see Fig. \ref{fig1}(c)], as can be shown using $S^2=\mathbb{1}$ \cite{Muller:kz,Terhal:2015ks}. Note that this operation differs from the one given in Eq.~\eqref{eq2} by an additional swap, $U_{\rm step}=(\mathbb{1}\otimes S)U_{\rm step}'$. This swap is convenient since it guarantees 
 that in each stroboscopic step a new (unused) copy is coupled to the state that is propagated according to Eq.~\eqref{basic_step} while involving only neighboring copies (see Appendix \ref{App:Protocol}) .  

(iii) Measurement of the ancilla atom in the $Z$ basis, with outcomes $\pm1$.

The average over many runs gives $\mean{Z}_n$, and the eigenvalues of the density operator can be extracted via  a Fourier transform of Eq.~\eqref{Ramsey} over the simulated time $t_n$. Note that one can choose any subset of modes $\mathcal{A}$ by addressing the corresponding sites with the ac-Stark laser in step (ii-b). In particular, the addressability  allows measuring the spectrum of any reduced state $\rho_{\mathcal{A}}$, as well as the spectrum of the entire state $\rho$. 

We note that in systems with a conserved quantum number, such as the total number of atoms, it can be useful to access the ES in a quantum-number-resolved way \cite{Regnault:2009bf}. In fact, our protocol can be easily adjusted to measure the ES in different number sectors \cite{Lauchli:2013jga}, via a \textit{preselection} of the copies by measuring the total number of atoms in subsystem $\mathcal{B}$ in all $n$ copies before step (i). Beyond the conceptual asset of obtaining richer information  about the entanglement structure, this also has the advantage of increasing the spectral resolution, as discussed below and pointed out in calculations of ES from quantum Monte Carlo simulations \cite{Chung:2014bg}. 

We further point out that our protocol is not limited to the implementation of the controlled phase shift using an ancilla atom and the Rydberg blockade mechanism. Alternatively one can place the atoms in an optical cavity \cite{Zhang:2012gw,Landig:2015et}  and use different photon number states as ancillary system.  The  different ac-Stark shift experienced by the atoms allows us to implement the controlled phase gate \cite{Swingle:2016td}.

While in the above discussion we focussed on a one-dimensional setting, it is easy to see that the protocol equally applies to other configurations such as two-dimensional systems. Instead of preparing several copies in one-dimensional tubes, in step (i) the copies are, in this case, created in two-dimensional (2D) layers as indicated in Fig.~\ref{fig2}(d). All other steps are identical to the ones described above. Experimentally, site-resolved addressing is more challenging in this 2D setting, as it involves also layer-resolved addressing. This could be achieved, for example, by combining quantum gas microscopes with individual addressing techniques using magnetic-field gradients \cite{Daley:2008ih}. Probing the ES in such 2D systems would allow us to diagnose, e.g., topological order in realizations of fractional quantum Hall states with cold atoms \cite{Sorensen2005,Hafezi2007}.

\begin{figure}
\includegraphics[width=1\columnwidth]{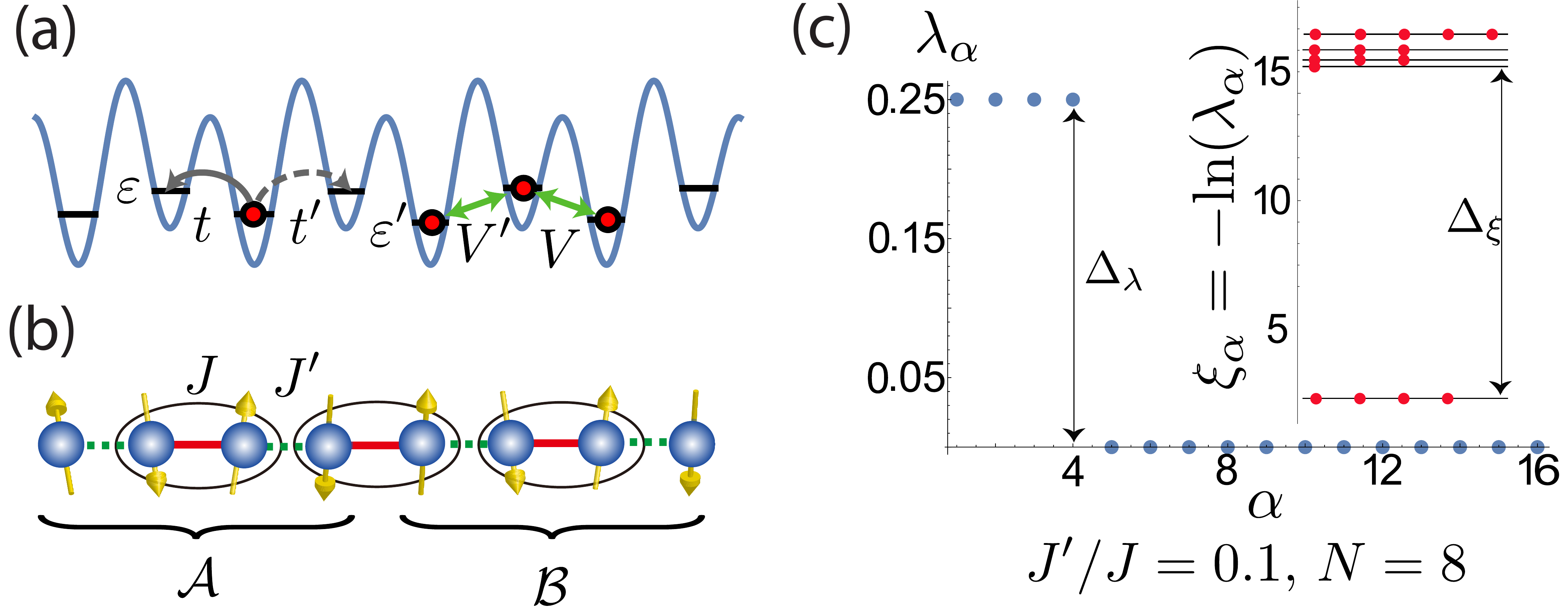}
\caption{Entanglement spectrum of the extended 1D Hubbard model supporting symmetry-protected topological order in the ground state. (a) Optical lattice with a two-site unit cell corresponding to the Hamiltonian \eqref{ExtendedBH}. The Hubbard parameter alternate between even and odd sites $t_{2j}\equiv t$, $t_{2j-1}\equiv t'$, and analogously for $\varepsilon_{j}$ and $V_j$. (b) In the hard-core boson regime, this can be mapped to an alternating-bond  spin-1/2 Heisenberg model \eqref{HamiltonianJ1J2}, with $J=t/2=V/4$ and $J'=t'/2=V'/4$ (see Appendix \ref{mapping}).  The strong antiferromagnetic bond (red) favors the formation of a spin-singlet dimer in the small $J'/J$ limit. (c)~ES  $\lambda_\alpha$ of a bipartite split of regions $\mathcal{A}$ and $\mathcal{B}$, and corresponding entanglement energies $\xi_\alpha=-\text{ln}(\lambda_\alpha)$  (inset) ($J'/J=0.1$, on $N=8$ sites), for one of the four degenerate ground states of Eq.~\eqref{HamiltonianJ1J2}.}
\label{J1-J2ES}
\end{figure}

\section{Illustration for the Haldane chain}
In the following we illustrate this protocol on the example of a Hubbard model. Here, we focus on an analysis of the simplest model in one dimension with symmetry-protected topological (SPT) order \cite{ChenScience} and show that the protocol presented in Sec.~\ref{Implementation} allows us to determine the largest eigenvalues of the ES, especially its gap and topological degeneracies. In particular, we consider an extended Bose-Hubbard model with nearest-neighbor interactions:
\begin{align}
\nonumber H & = - \sum_{j} t_j (a^\dag_{j} a_{j+1}+a^\dag_{j+1} a_{j})+\varepsilon_j n_j +\no\\
&+\sum_{j} U n_j (n_j-1) + V_j n_j n_{j+1}, \label{ExtendedBH}
 \end{align}
 where $n_j=a_j^\dag a_j$. 
As illustrated in Fig.~\ref{J1-J2ES}(a), we consider a superlattice with a two-site unit cell with the hopping amplitude $t_j$ and the on-site potential $\varepsilon_j$. The interacting part of the Hamiltonian contains the standard on-site interaction with strength $U$ and nearest-neighbor terms with strength $V_j$. Such extended Hubbard models have been realized recently in experiments with cold atoms with off-site interactions  stemming from magnetic dipole-dipole interactions \cite{Baier:2016ga}. The ground-state phase diagram of the Hamiltonian \eqref{ExtendedBH} can support a topologically nontrivial Haldane phase with a nonlocal string order parameter \cite{HIDA:1992bq,Tzeng:2015ut}. This can be most easily seen in the ``hard-core" boson limit ($U\rightarrow \infty$), where at most one boson can occupy a single lattice site. In this limit the problem can be  mapped into a spin-$\frac{1}{2}$ chain by $a^\dag_j \rightarrow (-1)^j (\sigma_j^{x}+i\sigma_j^{y})/2$ (with $\vec{\sigma}_j$ the Pauli operators). For a  proper choice of the Hubbard parameters (cf.~Appendix \ref{mapping}), the Hamiltonian \eqref{ExtendedBH} can be mapped into the alternating-bond Heisenberg chain \cite{Cross:1979hl, HIDA:1992bq, Tzeng:2015ut} 
\begin{align}
\label{HamiltonianJ1J2}
H_\text{eff} =  \sum_{j=1}^{N/2}  [ J (\vec{\sigma}_{2j} \cdot \vec{\sigma}_{2j +1}) + J'  (\vec{\sigma}_{2j-1}  \cdot \vec{\sigma}_{2j })].
\end{align}
In the case of $J, J' > 0$, the ground state of Eq.~\eqref{HamiltonianJ1J2} displays alternating strong-weak antiferromagnetic (AF)  bonds in the chain, as illustrated in Fig.~\ref{J1-J2ES}(b) by solid and dashed bonds.  In the limit of vanishing coupling on every second bond $(J'=0)$, the spins dimerize in spin singlets [indicated by ellipses in Fig.~\ref{J1-J2ES}(b)], while the two spins on the end of the chain are decoupled and hence form free edge states. This leads to a fourfold degeneracy of the ground state in an open chain, which survives in the thermodynamic limit also for finite $J'$, where the system is analogous to the spin-1 AKLT model \cite{Affleck:1987jy}. 

The Haldane phase in this regime falls into the category of SPT order and is protected by dihedral symmetries, time-reversal symmetry and inversion symmetry \cite{HIDA:1992bq}. Its topological properties can be probed by a direct measurement of the nonlocal string order parameter \cite{HIDA:1992bq,Tzeng:2015ut}. Here we use the Haldane chain as an example to illustrate how our protocol reveals topological order in terms of the ES, which is a more generic detection tool, as it also applies to situations where no such string order parameter exists. The ES in the ground state shows a fourfold degeneracy if the chain is bipartitioned along a $J$ bond \cite{Tzeng:2015ut} [cf.~Fig.~\ref{J1-J2ES}(c)]: If the size of subsystem $\mathcal{A}$ is larger than the edge-state correlation length, $\rho_\mathcal{A}$ can effectively be decomposed into two parts, $\rho_\mathcal{A} = \rho_{\textrm{edge}, \mathcal{A}} \otimes \rho_{\text{bulk}, \mathcal{A}}$, where $\rho_{\textrm{edge}, \mathcal{A}}$ describes the state of the edge mode in subsystem $\mathcal{A}$ (i.e., on the left in Fig.~\ref{J1-J2ES}) and $\rho_{\text{bulk}, \mathcal{A}}$ the state of the bulk modes in subsystem $\mathcal{A}$ \cite{ChenScience}.
The edge mode $\rho_{\textrm{edge}, \mathcal{A}}$ contributes a (trivial) factor of 2 to the degeneracy in the ES, while another factor of 2 stems from the entanglement in the bulk wave function $\rho_{\textrm{bulk}, \mathcal{A}}$.  
This degeneracy can be seen as a simple illustration of the bulk-edge correspondence in the ES \cite{Chandran:2011ex, Qi:2012jv}:  The bipartition between $\mathcal{A}$ and $\mathcal{B}$ effectively frees edge spin(s) in $\mathcal{A}$ and forms an ``entanglement edge mode"  [Fig.~\ref{J1-J2ES}(b)]. The gap in the ES, remaining finite in the thermodynamic limit, separates the physical and entanglement edge modes from bulk excitations and shrinks when the correlation length increases with $J'/J$. 

\begin{figure}
\includegraphics[width=0.9\columnwidth]{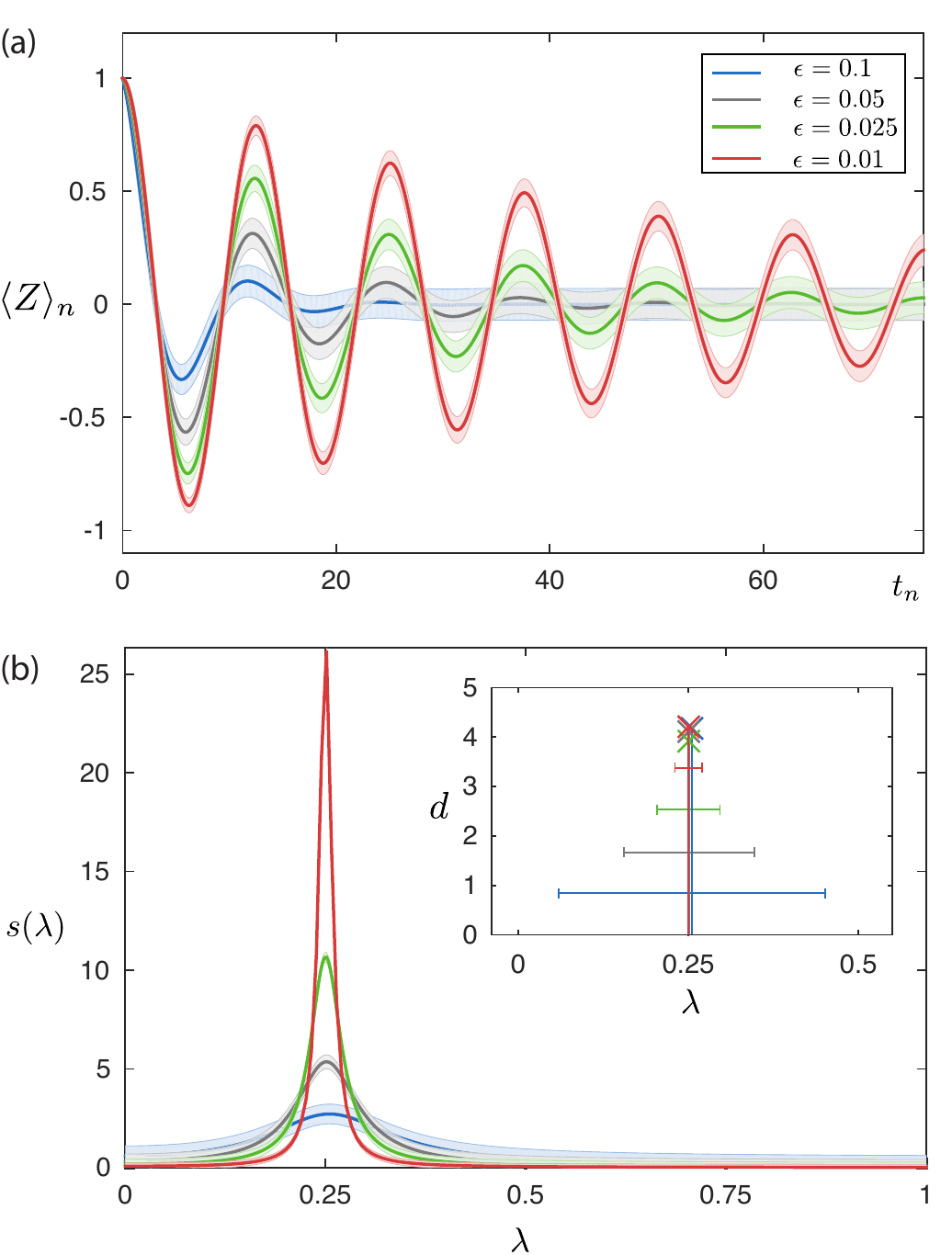}
\caption{Analysis of the protocol. (a)~Ramsey signal as a function of the stroboscopic evolution time $t_n$ for a density operator with four degenerate eigenvalues $\lambda_\alpha=1/4$, $(\alpha=1,\dots, 4$), using different sizes of the time step $\epsilon$. The finite size of the time step leads to a damping of the signal over a time $\tau\sim\mathcal{O}(\epsilon^{-1})$ (cf. Appendix \ref{App:Protocol}). We indicate the measurement uncertainty due to shot noise $\Delta Z_n=\sqrt{(1-\mean{Z}_n^2)/N_{\rm shot}}$ for $N_{\rm shot}=200$ measurements per point. (b)~Fourier-transformed Ramsey signal $s(\nu)=\sum_n \epsilon \mean{Z}_n\cos(2t_n \nu)$, with the uncertainty $\Delta s\sim\epsilon\sqrt{\sum_n \Delta Z_n^2}$. Each eigenvalue of the density operator gives rise to a Lorentzian whose width is determined by the size of the stroboscopic step. The product of the height and the width of the peak reveals that it stems from $d=4$ eigenvalues at $\lambda= 0.25$. This is depicted in the inset by the circles, where the horizontal bars correspond to the width of the peak and indicate the spectral resolution. To observe the signature of the fourfold degeneracy around $n_{\rm max}\sim 150$ copies are necessary ($\epsilon=0.1$). The position of the maximum, together with the height and the width of the resonance, allows one to determine the weight of the peak, i.e., the degeneracy $(d)$ of the corresponding eigenvalue  (see text). The inset shows the position of the resonance and its degeneracy $d$ for the peak at $\lambda\approx0.25$ for the different values of $\epsilon=0.1,0.05,0.025,0.01$. In all cases, one finds $d\sim 4$, as expected for fourfold degenerate eigenvalues. The horizontal bar indicates the width of the peak, $\sigma_\alpha$. }
\label{Fig4}
\end{figure}

In Fig.~\ref{Fig4}, we illustrate how our protocol would measure the ES and detect both the degeneracy of the ground state and the degeneracy of the reduced system. 
We plot in Fig.~\ref{Fig4}(a) the evolution of $\mean{Z}_n$ as a function of used copies, for different values of the stroboscopic step size $\epsilon$, applied to a system described by a mixture of the four degenerate ground states of the Haldane chain [with $J'/J=0.1$, as in Fig.~\ref{J1-J2ES}(c)], i.e., a thermal state at a temperature well below the excitation gap. As shown in Fig.~\ref{J1-J2ES}(c), the ES is dominated by the fourfold-degenerate largest eigenvalue $\lambda\approx1/4$. The Ramsey signal $\mean{Z}_n$ shows the corresponding characteristic oscillations as a function of the used copies $n$. The decay of the signal, which is not captured by Eq.~\eqref{Ramsey}, is due to the higher-order contributions for a finite step size $\epsilon$ (see Appendix \ref{Trotter_errors} for an analytical formula). 
In Fig.~\ref{Fig4}(b), we show the Fourier transform of this signal, $s(\nu)=\sum_n \epsilon \mean{Z}_n\cos(2t_n \nu)$. Because of the decay of the signal, it does not display sharp peaks at the eigenvalues of the density operator, but each eigenvalue $\lambda_{\alpha}$ gives rise to a Lorentzian profile (with width $\sigma_\alpha\sim\epsilon$), centered around it, $s(\nu)\simeq\sum_{\alpha}\frac{\lambda_\alpha}{4}\frac{\sigma_\alpha}{(\nu-\lambda_\alpha)^2+\sigma_{\alpha}^2}$. Since the eigenvalues appear as resonance frequencies, but also as the weight of the peak, one can determine the degeneracy $d$ of an eigenvalue by the product of the height and the width of the corresponding peak i.e., the degeneracy $d=4s(\lambda_\alpha)\sigma_\alpha$. This product is shown in the inset of Fig.~\ref{Fig4}(b), revealing the fourfold degeneracy of $\lambda=0.25$. Note that, for a time step $\epsilon=0.1$, this degeneracy is already visible for a propagation time $t_{n_{\rm max}}\sim 15$, i.e., $n_{\rm max}\sim 150$ copies. 

\section{Spectral resolution}
In general, the spectral resolution $\delta \lambda$ that is  achievable with this protocol is determined by the total evolution time [$\delta\lambda \sim 1/(\epsilon n_{\rm{max}})$]. For a given $n_{\rm max}$, the optimal step size scales as $\epsilon\sim 1/\sqrt{n_{\rm max}}$, maximizing the total integration time $t_{n_{\rm max}}$, while minimizing the decay of the Ramsey signal due to Trotter errors (see Appendix \ref{Trotter_errors}). The corresponding spectral resolution scales as $\delta\lambda\sim 1/\sqrt{n_{\rm max}}$.
According to the Nyquist sampling theorem the signal $\mean{Z}_n$ needs to be measured only at times $t_n$ with $n\simeq m/(2\epsilon)$ ($m=1,2,\dots$), such that the total number of measurement points scales as $\sim 1/(2\delta \lambda)$. Note that beyond the Fourier analysis illustrated above, there exist powerful numerical techniques, based on Prony's algorithm that are tailored to extract frequencies of such damped exponentials \cite{Barkhuijsen:1987kr}. 
Moreover, note that shot noise is not a severe limiting factor in our scheme. The corresponding uncertainty can be bounded by $\Delta Z_n=\sqrt{(1-\mean{Z}_n^2)/N_\text{shot}}$ and $\Delta s \lesssim \epsilon\sqrt{n_{\rm max}/ N_\text{shot}}$ respectively. Thus, the signal-to-noise ratio for the smallest resolvable peak (at $\lambda\sim\epsilon$) can be estimated as $\Delta s/s\sim 1/\sqrt{N_{\rm shot}}$.  This is in striking contrast to other approaches used to measure the ES based on measurements of Renyi entropies of different orders \cite{Daley:2012ud,Song:2012cp,Chung:2014bg}. In these approaches, the spectrum is determined by the roots of the characteristic polynomial, whose coefficients can be expressed in terms of the moments of the density operator via Newton's identities. One of the main difficulties in such an approach is that root-finding algorithms can be extremely sensitive to noise in the coefficients of the characteristic polynomial, i.e., shot noise \cite{Wilkinson:1959uw}. 

\section{Conclusion  and Outlook} In this work, we showed how one can access the spectrum of an arbitrary (reduced) density operator of a many-body system by implementing a hardware protocol, instead of full quantum state tomography.  While here we proposed a interferometric scheme that is within technological reach with upcoming cold-atom experiments, the scheme could be generalized to other quantum simulation platforms such as ion trap and circuit-QED architectures. Moreover, one can envisage it as the main building block of a full quantum principal component algorithm \cite{Lloyd:2014gc}, where instead of a single Rydberg atom several ancilla qubits control the protocol, and the spectrum could be extracted via quantum phase estimation \cite{Nielsen:2011:QCQ:1972505}. While here our analysis focused on bosonic atoms, we expect that analogous methods can also be employed for fermionic species \cite{Pichler:2013bs}. Finally, we want to point out that the concept of density matrix exponentiation discussed here has applications beyond spectral estimation as explored recently in Refs.~\cite{Marvian:2016uni,Kimmel:2016ham}. 

\section{Acknowledgements} We thank L.~Bonnes, E.~Kapit, A.~M.~L\"auchli, M.~Ozols, A.~Sterdyniak, A.~Turner, and F.~Verstraete for useful discussions. H.~P. was supported
by the NSF through a grant for the Institute for Theoretical Atomic, Molecular, and Optical Physics at Harvard
University and the Smithsonian Astrophysical Observatory. The work at Innsbruck is supported by the ERC Synergy Grant UQUAM, the Austrian Science Fund through SFB FOQUS, and the EU FET Proactive Initiative SIQS and RYQS. The work at the University of Maryland was supported by ONR-YIP, ARO-MURI, AFOSR, NSF-PFC  at the Joint Quantum Institute, and the Sloan Foundation. Kavli Institute for Theoretical Physics is supported by NSF PHY11-25915.

H.P., G.Z., and A.S. contributed equally to this work. 

 \begin{appendix}

\section{Protocol}\label{App:Protocol}
Here, we outline the derivation of Eq.~\eqref{Ramsey} of the main text and elaborate on the decomposition of a single stroboscopic step into processes that can be implemented in cold-atom experiments. 
\subsection{Ramsey interferometer}\label{app:Ramsey}
The unitary giving the Ramsey interferometer with $n$ steps described in the main text can be compactly written in the form
$
U_{\rm Ramsey}=\prod_{k=1}^{n}U_{\rm step}^{(k,k+1)},
$
where we imply an ordering of the product defined as $\prod_{k=1}^{n}A_k=A_n A_{n-1} \cdots A_1$. Also note that we indicate the copies on which the operators act in the superscript; e.g. $S^{(k,l)}$ interchanges the quantum states in copy $k$ and $l$ and leaves the other copies invariant. As in the main text we suppressed this whenever there is no danger of confusion to simplify notation. Here, we want to prove Eq.~\eqref{Ramsey}, which states
\begin{align}
\mean{Z}_n&=\tr{Z U_{\rm Ramsey}\lr{\ket{0}\bra{0} \otimes \rho^{\otimes (n+1)}} (U_{\rm Ramsey})^\dag}\no\\
&=\textrm{tr}\{\rho\cos(2t_n \rho)\}=\sum_{\alpha}\lambda_\alpha\cos(2t_n \lambda_\alpha).
\end{align}
To this end we first note that one can write 
$U_{\rm Ramsey}=\lr{\prod_{k=1}^n S^{(k,k+1)}}U_{\rm Ramsey}'$,
where we defined 
$
U_{\rm Ramsey}'\equiv\prod_{l=2}^{n+1} U_{\rm step}'^{(1,l)}
$
with 
\begin{align}\label{A1}
U_{\rm step}'^{(k,l)}&\equiv\ket{-}\bra{-}\otimes e^{i\epsilon S^{(k,l)}}+\ket{+}\bra{+}\otimes e^{-i\epsilon S^{(k,l)}}.
\end{align}
Since $S^{(k,k+1)}$ are unitary operators acting as the identity on the ancilla Hilbert space one can write the measurement result of the interferometer with $n$ steps, $\mean{Z}_n$, as 
\begin{align}
&\mean{Z}_n\!=\!\frac{1}{2}\Re\lr{\tr{\!\lr{\prod_{k=2}^{n+1} e^{i\epsilon S^{(1,k)}}\!}\!\rho^{\otimes (n+1)}\! \lr{\prod_{k=2}^{n+1} e^{-i\epsilon S^{(1,k)}}\!}^{\!\dag}\!}}.\label{AppendixRamsey}\no
\end{align}
Using the relation 
\begin{align}
\ptr{2}{e^{i\epsilon S} X\otimes Y e^{i\epsilon S}}&=\cos^2(\epsilon)X \tr{Y}-\sin^2(\epsilon)\tr{X}Y\no\\&+i\sin(\epsilon)\cos(\epsilon) (XY+YX),\no
\end{align}
one can calculate the trace by successively tracing out copy $k=2,3,\dots, n$ (in this order). 
This gives 
$\mean{Z}_n=\frac{1}{2}(\tr{M^n(\rho)}+\rm c.c.)$,
where we define the map
$
M(X)=(\cos(2\epsilon)+i\sin(2\epsilon)\rho)X+\sin^2(\epsilon)(X-\rho\tr{X}).
$
One can separate the leading-order contribution in the $\epsilon\ll1$ limit as
$
\tr{M^n(\rho)}=\tr{\rho(\cos(2\epsilon)+i\sin(2\epsilon)\rho)^n}+R_n(\rho,\epsilon),
$
where the remainder can be bounded as follows:
\begin{align}
|R_n(\rho,\epsilon)|\leq \sum_{k=1}^n 2^{k}\sin^{2k}(\epsilon)\binom{n}{k}=((1+2\sin^2(\epsilon))^n-1).\no
\end{align}
Here we used $|\tr{\rho(\cos(2\epsilon)+i\sin(2\epsilon)\rho)^k}|\leq 1$ for $k=0,1,2,\dots$.
For $\epsilon\rightarrow 0$ and $n$ such that $t_n=n\epsilon=\rm const$, we get
$|R_n(\rho,\epsilon)|\leq (e^{2\epsilon t_n}-1)\rightarrow 0$ for $\epsilon t_n\ll1$,
such that 
\begin{align}
\mean{Z}_n&=\frac{1}{2}(\tr{\rho(\cos(2\epsilon)+i\sin(2\epsilon)\rho)^n}+{\rm c.c.})+\mathcal{O}(\epsilon^2 n)\no\\
&=\sum_{\alpha}\lambda_\alpha\cos(2\lambda_\alpha t_n)+\mathcal{O}(\epsilon t_n).
\end{align}
To leading order, this is equal to Eq.~\eqref{Ramsey} given in the main text.

\subsection{Single stroboscopic step}\label{app:singlestep}

Here we explicitly show that the gate sequence described in the main text indeed gives rise to the Ramsey interferometer leading to Eq.~\eqref{Ramsey}. To this end, we first note the following identity that can be checked with straightforward algebra (using $S^2=\mathbb{1}$)
$
U_{\rm step}'^{(k,l)}=U_{\rm c-swap}^{(k,l)} \,U_\epsilon\, U_{\rm c-swap}^{(k,l)},
$
decomposing the operator \eqref{eq2} into a rotation of the ancilla, $U_\epsilon$, given in Eq.~\eqref{Usinglequbit} and the controlled swap $U_{\rm c-swap}^{(k,l)}$ (also called Fredkin gate) which exchanges the quantum state in copies $k$ and $l$ based on the state of the ancilla qubit as
\begin{align}\label{Fredkin}
U_{\rm c-swap}^{(k,l)}=\ket{1}\bra{1}\otimes \mathbb{1}+\ket{0}\bra{0}\otimes S^{(k,l)}.
\end{align}

Next we decompose the Fredkin gate into the three operations (i)-(iii) given in the main text. 
To this end we recall that for bosons the local swap operator, $S_{j}^{(k,l)}$ interchanging the state of the atoms on site $j$ of the two involved copies acts as $S_j^{(k,l)}: a_{j,k}\leftrightarrow a_{j,l}$ \cite{Daley:2012ud}. The global swap $S^{(k,l)}$ is a product these local swaps $S^{(k,l)}=\prod_{j=1}^M S_j^{(k,l)}$. Analogously, the swap operator in subsystem $\mathcal{A}$ is given by $S_{\mathcal{A}}^{(k,l)}=\prod_{j\in \mathcal{A}}S_j^{(k,l)}$.
Each of the local swaps can be written as $S_j^{(k,l)}=(\tilde U_{\textrm{BS}}^{(k,l)})^\dag(-1)^{a_{j,k}^{\dag}a_{j,k}}\tilde U_{\textrm{BS}}^{(k,l)}$ where $\tilde U_{\textrm{BS}}^{(k,l)}$ is the unitary that corresponds to the so-called beam splitter, which maps the symmetric (antisymmetric) modes as $(\tilde U_{\textrm{BS}}^{(k,l)})^\dag a_{j,k}\tilde U_{\textrm{BS}}^{(k,l)}=(a_{j,k}-a_{j,l})/\sqrt{2}$ and $(\tilde U_{\textrm{BS}}^{(k,l)})^\dag a_{j,l}\tilde U_{\textrm{BS}}^{(k,l)}=(a_{j,k}+a_{j,l})/\sqrt{2}$. 
With this (and using the fact that $\tilde U_{\rm BS}^{(k,l)}$ acts as the identity on the ancilla Hilbert space) one can write the Fredkin gate as
$
U_{\rm c-swap}^{(k,l)}=(\tilde U_{\textrm{BS}}^{(k,l)})^\dag U_{\rm c-phase}^{(k)}\tilde U_{\textrm{BS}}^{(k,l)},
$
where 
\begin{align}
U_{\rm c-phase}^{(k)}=\ket{1}\bra{1}\otimes \mathbb{1}+\ket{0}\bra{0}\otimes (-1)^{\sum_j a_{j,k}^{\dag}a_{j,k}}
\end{align}
is the gate given in Eq.~\eqref{cPhasegate}, such that
\begin{align}
U_{\rm step}'^{(k,l)}=(\tilde U_{\textrm{BS}}^{(k,l)})^\dag U_{\rm c-phase}^{(k)} \,U_\epsilon\,  U_{\rm c-phase}^{(k)}\tilde U_{\textrm{BS}}^{(k,l)},
\end{align}
where we used $[\tilde U_{\rm BS}^{(k,l)},U_\epsilon]=0$.

Noting that $\tilde U_{\rm BS}^2=S$ (up to an irrelevant phase) we obtain the gate sequence given in the main text
\begin{align}
U_{\rm step}^{(k,l)}&\equiv S^{(k,l)} U_{\rm step}'^{(k,l)}=\tilde U_{\textrm{BS}} U_{\rm c-phase} \,U_\epsilon\, \tilde U_{\rm c-phase}\tilde U_{\textrm{BS}}.\no
\end{align}
Finally, note that formally the operator $\tilde U_{\rm BS}$ differs form the beam splitter given in Eq.~\eqref{BS} by local phase shifts. However, these local phase shifts are irrelevant in the presence of an atom-number super-selection rule, such that we can simply make the replacement $\tilde U_{\rm BS}\rightarrow U_{\rm BS}$. 

\subsection{Exponential decay for finite $\epsilon$}\label{Trotter_errors}
The expressions given in Appendix \ref{app:Ramsey} can be used to calculate the measurement outcome for any $\epsilon$. However, it is instructive to consider a simpler circuit where the evolution with $U_S(\epsilon)$ is only applied only to one arm of the interferometer, while in the other arm of the interferometer the state is left unchanged. One can straightforwardly  adjust the protocol to such a setting by using a third internal level for the ancilla system. Such a modified scheme is more convenient for an analysis of effects of a finite $\epsilon$ since it allows for an analytical calculation of the measurement result to all orders in $\epsilon$. 
This modified Ramsey interferometer will give a measurement result $\mean{Z}_n=\Re\{E(n,\epsilon)\}$, with 
\begin{align}
E(n,\epsilon)=\textrm{tr}\{U_{S}^{(1,2)}(\epsilon)U_{S}^{(1,3)}(\epsilon)\cdots U_{S}^{(1,n)}(\epsilon)\rho^{\otimes (n+1)}\},\no
\end{align}
where $U_{S,(a,b)}(\epsilon)=e^{-i\epsilon S^{(a,b)}}$. Since the products of swaps can be written as a cyclic permutation, one can use identities given in Refs.~\cite{Ekert:2002uu,Daley:2012ud} to express this in terms of moments of the density operator:
\begin{align}
E(n,\epsilon)&=\sum_{k=0}^n \left(\begin{array}{c}n \\k\end{array}\right)(-i)^k \cos^{n-k}(\epsilon)\sin^k(\epsilon)\textrm{tr}\{\rho^{k+1}\}\no\\
&=\sum_{\alpha}\lambda_\alpha (\cos(\epsilon)-i\lambda_\alpha\sin(\epsilon))^n\no.
\end{align}
One can easily note that this is a sum of damped exponentials, 
\begin{align}
\label{eq:evolve}
E(n,\epsilon)&=\sum_{\alpha}\lambda_\alpha e^{-\sigma_\alpha n-i\phi_\alpha n},
\end{align}
with $\sigma_\alpha=-\frac{1}{2}\log(\cos^2(\epsilon)+\lambda_\alpha^2\sin^2(\epsilon))\geq 0$ and $\phi_\alpha=-\arg\left(\frac{\cos(\epsilon)-i\lambda_\alpha\sin(\epsilon)}{(\cos^2(\epsilon)+\lambda_\alpha^2\sin^2(\epsilon))^{1/2}}\right)$.
A finite $\epsilon$ thus simply gives rise to a damped signal and a renormalization of the oscillation frequencies. Using Prony's analysis \cite{Barkhuijsen:1987kr}, one can extract the $\phi_\alpha$'s and the $\sigma_{\alpha}$'s from the measurement record, which in turn determine $\lambda_{\alpha}$ and its multiplicity from 
$
\lambda_\alpha=\sqrt{\frac{e^{-2\sigma_\alpha}-\cos^2(\epsilon)}{\sin^2(\epsilon)
}}
$
and
$
\lambda_\alpha=\frac{\tan(\phi_\alpha)}{\tan(\epsilon)}$. While these identities giving damped exponentials hold only when  the stroboscopic evolution is applied just to one arm of the interferometer, we found (numerically) similar behavior in the two-sided version presented in the main text, when $\epsilon$ is small. In particular, for the example studied in the main text the two versions are identical. 

\section{Experimental imperfections}\label{app:Imperfections}
\begin{figure}
\includegraphics[width=\columnwidth]{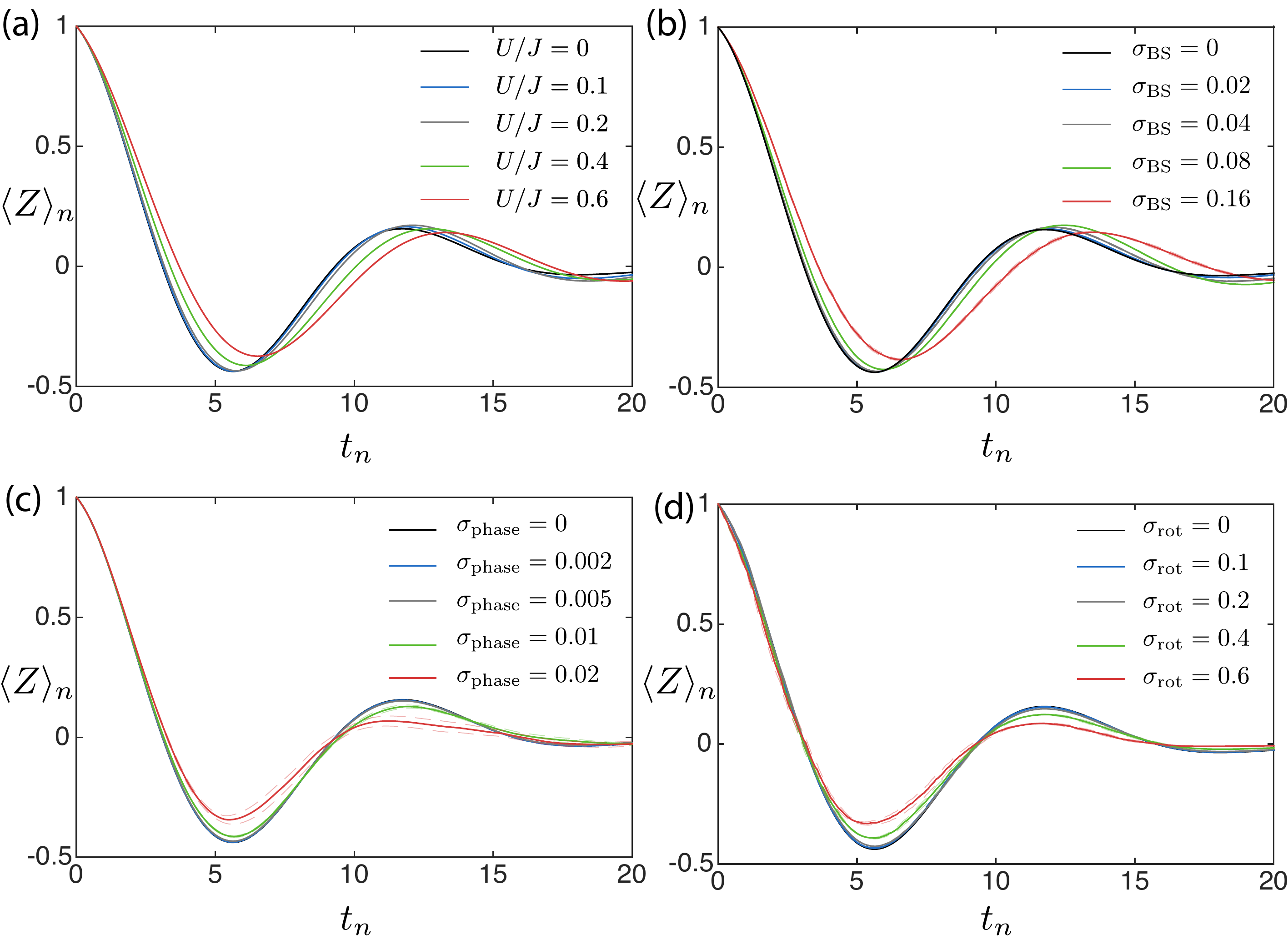}
\caption{Effects of imperfect implementations of the individual operations on the Ramsey signal, for the same system parameters as in Figs.~\ref{J1-J2ES} and \ref{Fig4} (with $N=4$ sites in each copy). We separately show the effects of (a)~residual interactions in the beam-splitter steps, (b) uncertainty of the tunnel times in the beam-splitter steps, (c) fluctuations in the phase acquired during the controlled phase gates, and (d) fluctuations in the rotation angle for single-qubit rotations (see text). For (b) and (c) we perform an average over different error realizations. The statistical uncertainties are indicated by the dashed lines. The size of the stroboscopic time step is $\epsilon=0.07$ in all cases.}
\label{Fig5}
\end{figure}

In this section we analyze the robustness of the protocol with respect to errors and imperfect implementations of the individual steps of the protocol. We consider imperfect beam splitters due to (i) residual on-site interactions $U$ (with $U/J\ll1$), i.e.,  
\begin{align}\label{BS_res_int}
U_{\rm BS}&\rightarrow\exp\bigg(i\frac{\pi}{4J}\sum_{j} \Big(J(a_{j,k}^\dag a_{j,k+1}+a_{j,k+1}^\dag a_{j,k})\no\\&+\frac{U}{2}(a_{j,k}^\dag a_{j,k}^\dag a_{j,k} a_{j,k}+a_{j,k+1}^\dag a_{j,k+1}^\dag a_{j,k+1} a_{j,k+1})\Big)\bigg),
\end{align} 
and (ii) imperfect timing $t_{\rm BS}\rightarrow\pi/(4J)(1+\xi_{
\rm BS})$, where $\xi_{\rm BS}\ll1$ quantifies the deviation from the ideal case.
Similarly, we model errors of the controlled phase shift by an error in the (controlled) phase, i.e., $t_{\rm phase}\rightarrow\pi\Delta/\Omega^2(1+
\xi_{\rm phase})$, and the error in the rotations of the ancilla qubit by $\epsilon\rightarrow \epsilon(1+\xi_{\rm rot})$. In all cases, we model $\xi_{\rm BS}$, $\xi_{\rm phase}$, and  $\xi_{\rm rot}$ as random variables with zero mean and standard deviation $\sigma_{\rm BS}$, $\sigma_{\rm phase}$, and $\sigma_{\rm rot}$. The corresponding Ramsey signal for all four cases is shown in Fig.~\ref{Fig5}. The calculations are done for the same parameters as in Fig.~\ref{J1-J2ES}; however, because of the large Hilbert space, we show only results for small systems of four lattice sites in each copy. For such small systems, the residual on-site interactions do not play a crucial role, as one can see in Fig.~\ref{Fig5}(a). This is not surprising as the number of particles in the subsystem is small. One expects that the deviation from the ideal case and the sensitivity to residual interactions increases with the number of particles, i.e., with the size of the subsystem. Among the other imperfections we find that our protocol is most sensitive to errors in the controlled phase gate. In the example shown in Fig.~\ref{Fig5}(c) the deviations from the ideal case are visible for fidelities $f_{
\rm phase}\equiv1-\sigma_{\rm phase}\lesssim0.99$. While such fidelities have not yet been achieved with Rydberg atoms \cite{Maller:2015is}, theoretical calculations indicate that they are within reach \cite{Xia:2013il}. Finally, random errors in the rotation angle of the ancilla qubit [Fig.~\ref{Fig5}(d)] can be simply interpreted as dephasing and lead to a decay of the Ramsey signal. They thus have a similar effect as Trotter errors because of a finite time step (see Appendix \ref{Trotter_errors}), and they eventually limit the maximal spectral resolution.

\section{Extended-Bose-Hubbard model and alternating-bond XXZ spin chain}\label{mapping}

We begin with the extended-Bose-Hubbard (EBH) model with alternating-strength nearest-neighbor hopping and interaction:
\begin{align}
\nonumber H = & \sum_{j=1}^{N/2} [(-t a^\dag_{2j} a_{2j+1}  - t'  a^\dag_{2j-1} a_{2j}+ \text{H.c.} )  \\
\nonumber & + V a^\dag_{2j} a_{2j} a^\dag_{2j+1} a_{2j+1}+ V' a^\dag_{2j-1} a_{2j-1} a^\dag_{2j} a_{2j}  \\
 & -\varepsilon a^\dag_{2j} a_{2j} - \varepsilon' a^\dag_{2j-1} a_{2j-1}  ]+ \sum_{j=1}^N  U a^\dag_{j} a^\dag_{j} a_{j}  a_{j}.
\end{align}
Here $t,t' > 0$ captures alternating nearest-neighbor hopping and $V,V'>0$ captures alternating nearest-neighbor repulsive interactions, while $U$ and $\varepsilon,\varepsilon'$ represent on-site repulsion and on-site potential energy respectively.
In the infinite-$U$ limit, i.e.~``hard-core" boson regime, the model we consider here can be mapped into a spin-$\frac{1}{2}$ chain, by setting $a^\dag_j \rightarrow (-1)^j \sigma^+_j$ and $a^\dag_j a_j \rightarrow \frac{1}{2} (\sigma^z_j + 1)$.      The definition of spin operators with an alternating sign (equivalent to a gauge transformation) is to remove the negative signs in front of $t$ and $t'$. At half filling (adjusting $\varepsilon= V$ and $\varepsilon'= V'$ ), one can get the alternating-bond XXZ model \cite{Cross:1979hl, HIDA:1992bq, Tzeng:2015ut} described by the following Hamiltonian:
\begin{align}\label{AppHamiltonianJ1J2}
\nonumber H_\text{eff} =&  \sum_{j=1}^{N/2}  [ J (\sigma^x_{2j} \sigma^x_{2j +1}+\sigma^y_{2j} \sigma^y_{2j +1}) + \frac{V}{4}  \sigma^z_{2j} \sigma^z_{2j +1} \\ 
&+  J' (\sigma^x_{2j-1} \sigma^x_{2j}+\sigma^y_{2j-1} \sigma^y_{2j }) + \frac{V'}{4}  \sigma^z_{2j-1} \sigma^z_{2j }],
\end{align}
where we have defined $J=\frac{1}{2} t$ and $J'=\frac{1}{2} t'$. Because of the open boundary conditions one needs an extra on-site potential in the first and last sites ($j=1$ and $j=N$) of $\delta \varepsilon=V/2$ for the mapping to be exact.
The Haldane phase and fourfold degeneracy in the ``ground state" of the ES exist in this model and can survive even in the limit of zero $zz$ coupling, i.e.~$V,V'=0$ \cite{Tzeng:2015ut}.  For simplicity, in the main text, we have chosen the off-site interactions to be $V=4J$ and $V'=4J'$ such that we realize an isotropic alternating-bond Heisenberg model.  However, we note that such fine-tuning is not essential for the observation of the degeneracy in the ES and its relation to topological order.

The ground states for $J'<0$ and $0<J'<J$ are in the same Haldane phase and can be adiabatically connected. At the quantum critical point $J=J'=1$, one recovers the Heisenberg model which leads to a gapless phase.  For $J' > J$, one enters to another Haldane phase with twofold entanglement degeneracy if we cut on the $J$ bond and a fourfold entanglement degeneracy if we cut on the $J'$ bond.

In the XXZ spin model the total magnetization along $S_z= \sum_j \sigma^z_j$ is a good quantum number, $[H_\text{eff} , S_z]=0$, corresponding to the conserved number of atoms in the extended Hubbard model, $N_{\rm tot}=\sum_j n_j$. 
Therefore, in the absence of external magnetic field, the fourfold degeneracy can be divided into different $S_z$ ($N_\text{tot}$) sectors. In the simple limit $J'=0$, the edge modes just involve the two spins in the end of the chain and are  composed of the four basis states $\ket{\uparrow}_L \ket{\uparrow}_R$, $\ket{\uparrow}_L \ket{\downarrow}_R$, $\ket{\downarrow}_L \ket{\uparrow}_R$ and $\ket{\downarrow}_L \ket{\downarrow}_R$. Here $L/R$ refer to the left/right edge mode at site $j=1$ and $j=N$, respectively. When mapped to bosons for the cold-atom realization, the four basis states become $\ket{1}_L \ket{1}_R$, $\ket{1}_L \ket{0}_R$, $\ket{0}_L \ket{1}_R$ and $\ket{0}_L \ket{0}_R$. The four states can obviously be divided into three different number sectors, with total boson number on the edges being $n=0$, $1$, and $2$.  While the $n=0$ and $n=2$ sectors are not degenerate, the $n=1$ sector is twofold degenerate. The same counting applies to the case when $J' \neq 0$ and the edge mode becomes extended.  Therefore, in the cold-atom experiments, if the total number of particles is fixed to $N/2$, the ground-state degeneracy is actually twofold at exactly half-filling. Nevertheless, in this case the entanglement degeneracy is still fourfold because of the contribution from both the physical and entanglement edge modes.

Since the SPT phase is short-range entangled, a disentangling into product states is always possible with a local unitary transformation or, equivalently, a finite-depth quantum circuit \cite{ChenScience}.  Therefore, as long as the system size is much larger than the correlation length, we are always allowed to decouple the edge and bulk states, i.e.~$\rho_\mathcal{A} = \rho_{\textrm{edge}, \mathcal{A}} \otimes \rho_{\text{bulk}, \mathcal{A}}$, after a local unitary transformation. 
\bibliographystyle{apsrev4-1.bst}



\end{appendix}
\end{document}